\pdfoutput=1

\documentclass[12pt]{JHEP3}

\let\ifpdf\relax
\usepackage{ifpdf}
\usepackage{multirow}
\usepackage{color}
\let\normalcolor\relax
\usepackage{mathdots}
\usepackage{mathtools}
\usepackage{graphicx}
\usepackage{cite}

\renewcommand{\bar}{\overline}

\definecolor{paper_blue}{rgb}{0.3,0.2,0.75}
\definecolor{paper_red}{rgb}{0.65,0.1,0.15}
\definecolor{paper_green}{rgb}{0.05,0.35,0.125}
\definecolor{paper_grey}{gray}{0.375}
\definecolor{perm}{rgb}{0.1,0.45,0.85}
\definecolor{deemph}{rgb}{0.7,0.7,0.7}

\newcommand{\eq}[1]{\begin{equation}#1\end{equation}}

\newcommand{\ab}[1]{\langle #1\rangle}
\newcommand{\newcap}{\mathrm{\raisebox{0.75pt}{{$\,\bigcap\,$}}}}
\newcommand{\mi}{{\rm\rule[2.4pt]{6pt}{0.65pt}}}
\newcommand{\pl}{\hspace{0.5pt}\text{{\small+}}\hspace{-0.5pt}}

\newcommand{\mathematica}[3]{\vspace{0.35cm}\noindent\boxed{\begin{minipage}{#1\textwidth}\begin{tabular}{lp{11cm}}{\color{paper_blue}{\scriptsize{\tt In[1]:}}\raisebox{-0.65pt}{{\scriptsize{\tt=}}}}&{\tt #2}\\{\color{paper_blue}{\scriptsize {\tt Out[1]:}}\raisebox{-0.65pt}{{\scriptsize{\tt=}}}}&{\tt #3}\end{tabular}\end{minipage}}\vspace{0.35cm}}

\definecolor{varcolor}{rgb}{0.1,0.55,0.25}
\definecolor{functioncolor}{rgb}{0.1,0.35,0.75}
\newcommand{\vardef}[1]{{\color{varcolor}{\sl #1}\rule[-1.05pt]{7.5pt}{.75pt}}}
\newcommand{\vardefms}[1]{{\color{varcolor}{\sl #1}\rule[-1.05pt]{15pt}{.75pt}}}
\newcommand{\vardefo}[1]{{\color{varcolor}{\sl #1}\rule[-1.05pt]{7.5pt}{.75pt}{\bf{\sl :}}}}
\newcommand{\vardefoo}[1]{{\color{varcolor}{\sl #1}\rule[-1.05pt]{15pt}{.75pt}{\bf{\sl :}}}}
\newcommand{\defn}[3]{~\\[-35pt]\begin{itemize}\item[]\indent\hspace{-21pt}$\bullet$\hspace{-.75pt} {\tt {\color{functioncolor}#1}\![}#2{\tt\,]\!:}#3\end{itemize}\vspace{-10pt}}
\newcommand{\defnNA}[3]{~\\[-35pt]\begin{itemize}\item[]\indent\hspace{-21pt}$\bullet$\hspace{-.75pt} {\tt {\color{functioncolor}#1}\!}#2{\tt\,\!:}#3\end{itemize}\vspace{-10pt}}
\newcommand{\defntb}[4]{~\\[-35pt]\begin{itemize}\item[]\indent\hspace{-21pt}$\bullet$\hspace{-.75pt} {\tt {\color{functioncolor}#1}\![}#2{\tt\,]\![}#3{\tt\,]\!:}#4\end{itemize}\vspace{-10pt}}
\newcommand{\var}[1]{{\tt{\color{varcolor}{\sl#1}}}}
\newcommand{\ind}{\hspace{4ex}}
\newcommand{\fun}[1]{{\color{functioncolor}#1}}

\title{{\LARGE Positroids, Plabic Graphs, \&\\\LARGE Scattering Amplitudes in {\sc Mathematica}}}
\author{Jacob L. Bourjaily\\Department of Physics, Harvard University, Cambridge, MA 02138}
\preprint{December 2012}

\abstract{The many intricate connections between scattering amplitudes, on-shell diagrams, and the positroid stratification of the Grassmannian has recently been described in detail. In order to facilitate the exploration of this rich correspondence, we have prepared a public {\sc Mathematica} package called ``{\tt positroids}'' which includes an array of useful tools including those for the construction of canonical coordinates for positroid configurations, the drawing of representative on-shell (plabic) graphs, and the evaluation of on-shell differential forms. This note documents the functions made available by the {\tt positroids} package; the package's source code together with a {\sc Mathematica} notebook containing many detailed examples of its functionality are included with this note's submission files on the {\tt arXiv}. 
}

\begin{document}
\maketitle

\tableofcontents


\section{Introduction}\label{introduction_section}
The recent work of \cite{ArkaniHamed:2012nw} presents a comprehensive summary of the extensive correspondence between ``on-shell diagrams'', \cite{Bern:1994cg,Bern:2007dw,Britto:2004nj,Drummond:2008bq,Cachazo:2008vp}, the Grassmannian contour integral described in\cite{ArkaniHamed:2009dn,Mason:2009qx,Bullimore:2009cb,Kaplan:2009mh,ArkaniHamed:2009dg}, scattering amplitudes in planar, maximally supersymmetric \mbox{(``$\mathcal{N}=4$'')} Yang-Mills (SYM), \cite{Britto:2004ap,Britto:2005fq,Cachazo:2008dx,Drummond:2008cr,ArkaniHamed:2008gz,Bourjaily:2010kw,ArkaniHamed:2010kv}, and the combinatorics and geometry of what is known as the {\it positroid} stratification of the Grassmannian, \cite{P,Williams:2003a}. At the heart of this story is the fact that scattering amplitudes can be represented (to all loop orders) in terms of on-shell diagrams, and that (reduced) on-shell diagrams can be fully characterized {\it combinatorially} by {\it permutations}---associated with left-right paths; moreover, these same permutations label the configurations of the positroid stratification of the Grassmannian $G(k,n)$ of $k$-planes in $n$ dimensions. These strata are naturally endowed with {\it positive} coordinates $\alpha_i$ and a canonical volume-form, \cite{FG3,FG4}, which when expressed in terms of positive coordinates, is simply: $d\log(\alpha_1)\wedge\!\cdots\!\wedge\!d\log(\alpha_d)$. Because on-shell diagrams can be {\it directly represented} (and computed) as integrals of this invariant volume-form over their corresponding positroid configurations, this makes the evaluation of on-shell diagrams exceedingly simple.

\newpage
We can illustrate this rich correspondence with the following example:~\\[-5pt]

\scalebox{.85}{\mbox{\hspace{0.45cm}\begin{minipage}[h]{\textwidth}\vspace{.2cm}
\eq{\hspace{-4.5cm}\begin{array}{ccccc}&&\text{$\hspace{-2cm}C(\alpha)\equiv${\footnotesize$\left(\!\begin{array}{@{}c@{}ccc@{}cc@{}c@{}cc@{}}1&\alpha_{8}&(\alpha_{5}\pl\,\alpha_{14} \alpha_{8})&\alpha_{5} \alpha_{11}&0&0&0&0&0\\
0&0&0&1&\alpha_{10}&(\alpha_{10} \alpha_{13}\pl\alpha_{4})&\alpha_{4} \alpha_{7}&0&0\\
\mi\,\alpha_{3} \alpha_{9}\,&0&0&0&0&0&1&\alpha_{6}&(\alpha_{3}\pl\,\alpha_{12} \alpha_{6})\\
\mi\,\alpha_{9}\,&0&\alpha_{1}&\alpha_{1} \alpha_{11}&0&\mi\,\alpha_{1} \alpha_{2}\,&\mi\,\alpha_{1} \alpha_{2} \alpha_{7}\,&0&1\end{array}\!\right)$}$\in G(4,9)\hspace{-2.5cm}$}
\\[-00pt]&\hspace{-.0cm}\raisebox{-5pt}{\includegraphics[scale=1]{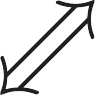}}\hspace{-2.05cm}&\hspace{-1.25cm}\raisebox{-00pt}{\includegraphics[scale=1]{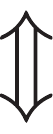}}\hspace{-1.5cm}&\hspace{-1.25cm}\raisebox{-5pt}{\includegraphics[scale=1]{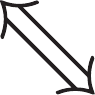}}\hspace{.0cm}\\[-45pt]\raisebox{-78pt}{\includegraphics[scale=1]{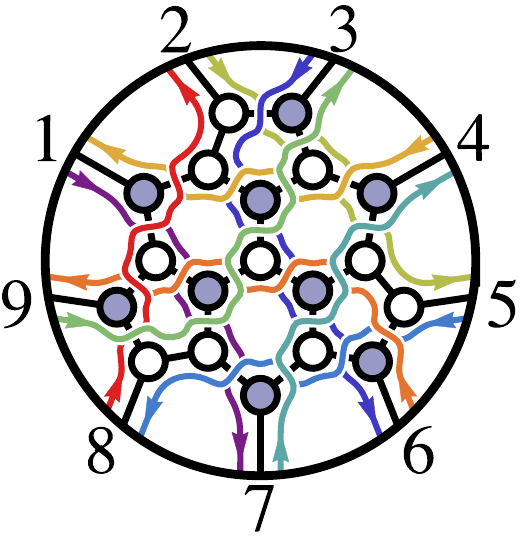}}\hspace{-0.75cm}&\raisebox{-5pt}{\includegraphics[scale=1]{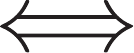}}\hspace{-2.75cm}&\hspace{-2cm}\begin{array}{c}\\[-30pt]\hspace{.25cm}\Bigg(\!\!\begin{array}{@{}ccccccccc@{}}\\\,\,1\,\,&\,\,2\,\,&\,\,3\,\,&\,\,4\,\,&\,\,5\,\,&\,\,6\,\,&\,\,7\,\,&\,\,8\,\,&\,\,9\,\,\\[-5pt]\downarrow&\downarrow&\downarrow&\downarrow&\downarrow&\downarrow&\downarrow&\downarrow&\downarrow\\[-2pt]7&5&6&1&8&9&4&2&3\\[-3pt]\!\!\!\!\!\!\!\!\!\!\!\!\!\!\sigma\!\equiv{\color{perm}\!\!\!\!\!\phantom{~,}\{7,\phantom{\}}\!\!\!\!\phantom{\equiv\sigma}}\!\!\!\!\!\!\!\!\!\!\!\!\!\!&{\color{perm}\phantom{,}5,}&{\color{perm}\phantom{,}6,}&{\color{perm}\!\!\phantom{,}10,\!\!}&{\color{perm}\phantom{,}8,}&{\color{perm}\phantom{,}9,}&{\color{perm}\!\!\phantom{,}13,\!\!}&{\color{perm}\!\!\phantom{,}11,\!\!}&{\color{perm}\!\!\!\phantom{\{}12\}\!\!\!}\end{array}\!\!\Bigg)\end{array}\hspace{-2cm} &\hspace{-2.05cm}\raisebox{-5pt}{\includegraphics[scale=1]{lr_arrow}}\hspace{-.1cm}&
\hspace{-0cm}\raisebox{-67pt}{\includegraphics[scale=1]{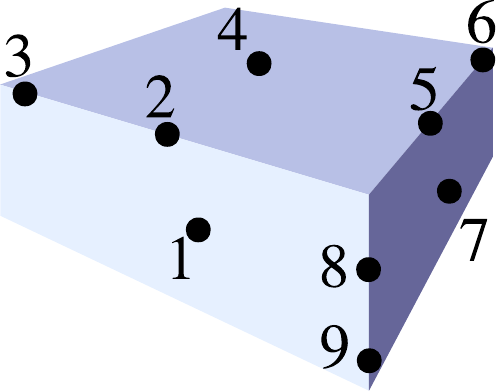}}\hspace{-0cm}
\\[-50pt]&\hspace{.0cm}\raisebox{-5pt}{\includegraphics[scale=1]{diag_arrow_2}}\hspace{-2.05cm}&\hspace{-1.25cm}\raisebox{-20pt}{\includegraphics[scale=1]{ud_arrow}}\hspace{-1.5cm}&\hspace{-1.25cm}\raisebox{-5pt}{\includegraphics[scale=1]{diag_arrow_1}}\hspace{.0cm}\\[-0pt]
&&
\displaystyle \hspace{-2cm}f_{\sigma}\!\equiv\!\!\int\!\!\frac{d\alpha_1}{\alpha_1}\!\wedge\!\cdots\!\wedge\!\frac{d\alpha_{14}}{\alpha_{14}}\,\delta^{k\times 4}\big(C(\alpha)\!\cdot\!\widetilde{\eta}\big)\delta^{k\times 2}\big(C(\alpha)\!\cdot\!\widetilde{\lambda}\big)\delta^{2\times (n-k)}\big(\lambda\!\cdot\!C(\alpha)^{\perp}\!\big)\phantom{}\hspace{-2cm}
\end{array}\nonumber\hspace{-4.5cm}}\end{minipage}\hspace{-1cm}}\vspace{.2cm}}~\\[5pt]

\noindent Starting with the on-shell diagram on the left, we find that it would be labeled (via left-right paths) by the permutation denoted $\sigma\!\equiv\!{\color{perm}\{7,5,6,10,8,9,13,11,12\}}$; this permutation {\it also} labels the Grassmannian configuration drawn (projectively\footnote{Here, each dot represents a column of the matrix $C(\alpha)$ viewed projectively as a point in $\mathbb{P}^3$.}) on the right---a configuration represented by the matrix $C(\alpha)$ above, parameterized by the positive coordinates $\alpha_i$; in terms of $C(\alpha)$, the corresponding on-shell `function' $f_\sigma$ associated with the diagram is determined by the integral at the bottom. 

We will not review these ideas here, but instead refer the interested reader to reference \cite{ArkaniHamed:2012nw} for a thorough introduction and summary together with a more comprehensive list of references to the existing literature. 

In order to help facilitate further investigation along these lines, however, we have prepared a {\sc Mathematica} package called ``{\tt positroids}''---which is documented in this note. The {\tt positroids} package makes available many of the essential tools required to investigate  the myriad connections between on-shell physics, scattering amplitudes, and the combinatorics and geometry of the positroid stratification of the Grassmannian.

\newpage 
\section{The {\sc Mathematica} Package {\tt positroids} }\label{using_positroids_section}
\subsection{Obtaining the {\tt positroids} Package and Demonstration Notebook}\label{obtaining_subsection}
From the abstract page for this paper on the {\tt arXiv}, follow the link ``other formats'' (below the option for ``PDF'') and download the ``source'' for the submission. The source-file will contain\footnote{Occasionally, the ``source'' file downloaded from the {\tt arXiv} is saved without any extension; this can be ameliorated by manually appending ``{\tt .tar.gz}'' to the name of the downloaded file.} the {\tt positroids} package's main source-code ({\tt positroids.m}), together with a demonstration notebook ({\tt positroids\rule[-1.05pt]{7.5pt}{.75pt}package\rule[-1.05pt]{7.5pt}{.75pt}demo.nb}) which describes with detailed examples many of the functions defined by the package.
\subsection{Using the {\tt positroids} Package }\label{using_subsection}

Upon obtaining the package, users should open and evaluate the {\sc Mathematica} notebook {\tt postroids\rule[-1.05pt]{7.5pt}{.75pt}package\rule[-1.05pt]{7.5pt}{.75pt}demo.nb}; this notebook will copy {\tt positroids.m} to the user's {\tt Application} directory for {\sc Mathematica}; this will allow {\tt positroids} to be started in any future notebook via the command:\\[5pt]
\mathematica{.8}{\raisebox{-2pt}{{\tt<<positroids.m}}}{\raisebox{-212pt}{\includegraphics[scale=.6]{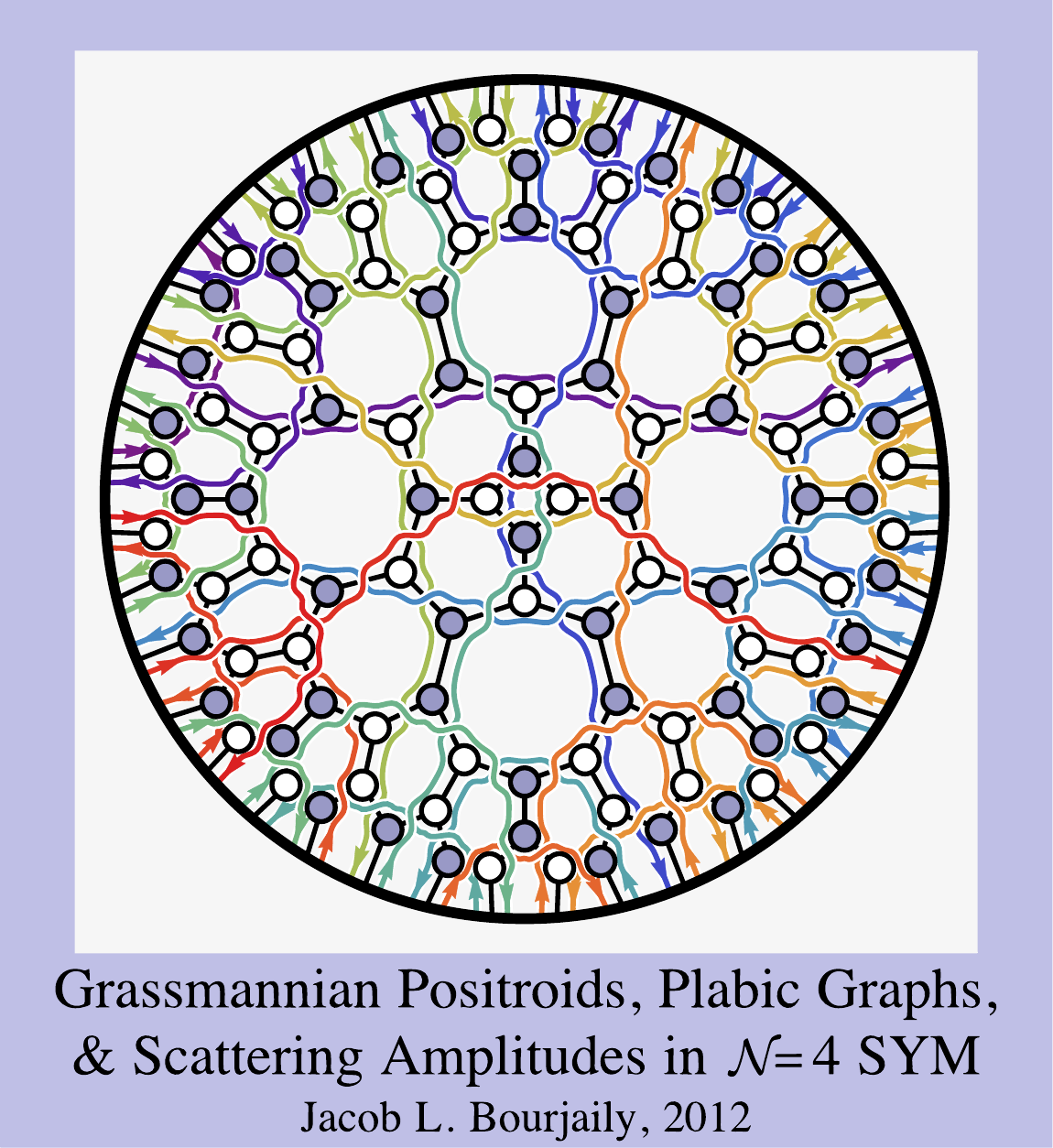}}}\\[-20pt]

\noindent (If the file ``{\tt positroids.m}'' has not yet beed copied to the user's {\tt Application} directory, then the package can be initialized by saving a notebook to the directory where {\tt positroids.m} is located, and evaluating ``{\tt SetDirectory[NotebookDirectory[]]}'' prior to the command ``{\tt <<positroids.m}''.)

\newpage

\section{Functions Defined by {\tt positroids}}\label{glossary_of_functions}

\subsection{Operations on Permutations Labeling Positroids}

\defn{boundary}{\vardef{permutation}}{returns a list of permutation labels for positroid cells in the boundary, $\partial$, of the cell labeled by \var{permutation}. For example, the co-dimension one boundaries of the generic configuration in $G_+(3,6)$ are: \\[5pt]
\mathematica{0.8}{boundary[\{4,5,6,7,8,9\}]}{{\tt \{\{5,4,6,7,8,9\},\{4,6,5,7,8,9\},\{4,5,7,6,8,9\},} ~\hspace{14cm}$~$ {\tt \{4,5,6,8,7,9\},\{4,5,6,7,9,8\},\{3,5,6,7,8,10\}\}}}\\[-20pt]
}

\defntb{checkOperator}{\vardef{permutation}}{\vardef{operator}}{returns:
\vspace{-.2cm}\eq{\left.\begin{array}{@{}ll}\bullet\;0\,&\text{if the minor ${\color{varcolor}(}$\var{operator}${\color{varcolor})}$ vanishes for the cell labeled by \var{permutation}};\\\bullet\;1&\text{if minor ${\color{varcolor}(}$\var{operator}${\color{varcolor})}$ is {\it non}-vanishing for the cell labeled by \var{permutation}.}\end{array}\right. \nonumber\vspace{-.2cm}}
\mathematica{.8}{\{checkOperator[\{3,5,6,7,8,10\}][\{1,2,3\}],\hspace{14cm}$~$\phantom{~}checkOperator[\{3,5,6,7,8,10\}][\{2,3,4\}]\}}{{\tt\{0,1\}}}\\[-20pt]
}

\defn{cyclicize}{\vardef{permutation}}{returns a {\it sorted} list of non-repeating permutations in the same cyclic class as \var{permutation}.\\[5pt]
\mathematica{0.8}{
cyclicize[\{6,5,8,7,10,9,12,11\}]}{
{\tt \{\{4,7,6,9,8,11,10,13\},\{6,5,8,7,10,9,12,11\}\}}}\\[-20pt]
}

\defn{cyclicRep}{\vardef{permutation}}{returns the lexicographically-first permutation in the cyclic-class of \var{permutation}.
}

\defn{decorate}{\vardef{permutation}}{takes an `ordinary' permutation $\sigma$ of $n$ integers, and returns a {\it decorated}, affine permutation $\widehat\sigma$, adding $n$ to the image of any $a$ such that $\sigma(a)<a$. For example, applying \fun{decorate} to the ordinary permutation associated with the on-shell diagram given in the introduction (\mbox{section \ref{introduction_section}}) gives: \\[5pt]
\mathematica{0.8}{decorate[\{7,5,6,1,8,9,4,2,3\}]\hspace{0pt}$~$}{{\tt\{7,5,6,10,8,9,13,11,12\}}}\\[-20pt]
}

\defn{dimension}{\vardef{permutation}}{gives the dimension of the positroid stratum labeled by \var{permutation}; for example,\\[5pt]
\mathematica{0.8}{dimension[\{3,5,7,6,8,14,10,12,11,13,16,21\}]\hspace{0pt}$~$}{{\tt20}}
}

\defn{dualGrassmannian}{\vardef{permutation}}{if \var{permutation} labels a $(2n\mi4)$-dimensional cell in the `momentum-space' Grassmannian $\widehat{C}\!\in\!G(k\pl2,n)$, then \fun{dualGrassmannian} returns the permutation label of the corresponding, $4k$-dimensional cell in the `momentum-twistor' Grassmannian $C\!\in\!G(k,n)$, and {\it vice-versa}; e.g., \\[5pt]
\mathematica{0.8}{dualGrassmannian[\{6,5,8,7,10,9,12,11\}]\hspace{14cm}$~$dualGrassmannian@\%}{{\tt\{2,5,4,7,6,9,8,11\}} ~\hspace{4cm}$~$ {\tt \{6,5,8,7,10,9,12,11\}}}
}

\defn{eulerCharacteristicTable}{\vardef{n},\vardef{k}}{returns a table indicating the numbers of $d$-dimensional cells in the positroid stratification of $G_+($\var{k},\var{n}$)$. (This data is generated using the combinatorial results of \cite{Williams:2003a}.) \\[5pt]
\mathematica{0.8}{eulerCharacteristicTable[10,5]}{\raisebox{-125pt}{$\begin{array}{c}\begin{array}{|c|r||l|c|}\hline\text{dim}&~\hspace{3cm}\#&\#\hspace{3cm}~&\text{dim}\\\hline24&1&10&25\\22&55&220&23\\20&715&2002&21\\18&4985&11240&19\\16&23210&44220&17\\14&78087&128100&15\\12&195315&276450&13\\10&362175&437112&11\\8&482670&482940&9\\6&432060&339360&7\\4&228102&126420&5\\2&54600&16800&3\\0&3150&252&1\\\hline&1865126&1865125&\\\hline\end{array}\\\mathbf{Euler~Characteristic}\!:\mathbf{1865126-1865125=1}\end{array}$}}
}

\defn{intersectionNumber}{\vardef{permutation},\vardefo{m}{\tt 4}}{returns the number of {\it isolated} solutions to \var{m}-dimensional kinematical constraints, $\Gamma^{\var{m}}(C)$, where $C$ is the positroid labeled by \var{permutation}. Supposing that $C$ is an $\!($\var{m}$\,k)$-dimensional cell in the momentum-twistor Grassmannian\footnote{\fun{intersectionNumber} also tests $(2n\mi4)$-dimensional cells in the momentum-space Grassmannian.}, \fun{intersectionNumber} counts the number of isolated points $C^*\!\in\!C\newcap Z^{\perp}$, where $Z$ is a {\it generic} configuration in $G($\var{m}$,n)$.\\[5pt]
\mathematica{.9}{\{intersectionNumber[\{6,5,8,7,10,9,12,11\}], \hspace{14cm}\phantom{~}intersectionNumber[\{15,14,8,7,21,20,19,13,12,26, \phantom{~~~~~~~~~~~~~~~~~~~~~}25,24,18,17,31,30,29,23,22,36\}]\}\hspace{0pt}}{{\tt\{2,34\}}}
}

\defn{inverseBoundary}{\vardef{permutation}}{returns a list of permutation labels for positroid cells in the {\it inverse}-boundary, $\partial^{-1}$, of the cell labeled by \var{permutation}.
}

\defn{legalPermQ}{\vardef{permutation}}{tests whether the list \var{permutation} does in fact denote a (possibly decorated) permutation on $n$ integers. 
}

\defn{necklace}{\vardef{permutation}}{if \var{permutation} labels a positroid configuration $C\!\in\!G(k,n)$, then \var{necklace} returns a list of $n$, $k$-tuples $A^{(a)}\equiv(A^{(a)}_1,\ldots,A^{(a)}_k)$ denoting the lexicographically-{\it minimal} non-vanishing minors starting from each column $a$. Recall from \cite{ArkaniHamed:2012nw,P} how the necklace encodes the list of all ranks of consecutive chains of columns of $C$; for example, the necklace for the configuration in $G(4,8)$ labeled by the permutation ${\color{perm}\{3,7,6,10,9,8,13,12\}}$ would be given by,
\vspace{-.3cm}\eq{\begin{array}{c@{$\!\!\!$}c}\raisebox{-70.5pt}{\includegraphics[scale=.9]{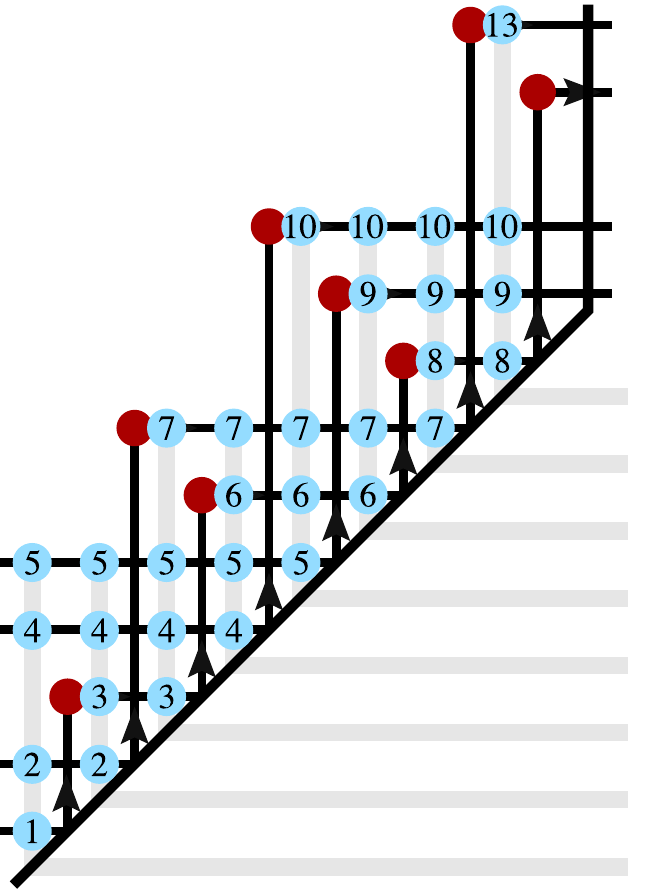}}&\begin{array}{l}~\\[-05.5pt]A^{(8)}=(8\,9\,10\,13)\\[1.8pt]A^{(7)}=(7\,8\,9\,10)\\[1.8pt]A^{(6)}=(6\,7\,9\,10)\\[1.8pt]A^{(5)}=(5\,6\,7\,10)\\[1.8pt]A^{(4)}=(4\,5\,6\,7)\\[1.8pt]A^{(3)}=(3\,4\,5\,7)\\[1.8pt]A^{(2)}=(2\,3\,4\,5)\\[1.8pt]A^{(1)}=(1\,2\,4\,5)\end{array}\\[10pt]\end{array}\vspace{-.3cm}}
This data is generated by the function \fun{necklace} according to:\\[5pt]
\mathematica{0.8}{necklace[\{3,7,6,10,9,8,13,12\}]\hspace{0pt}$~$}{{\tt\{\{1,2,4,5\},\{2,3,4,5\},\{3,4,5,7\},\{4,5,6,7\},\hspace{14cm}\phantom{1}\{5,6,7,2\},\{6,7,1,2\},\{7,8,1,2\},\{8,1,2,5\}\}}}
}


\defn{necklaceR}{\vardef{permutation}}{if \var{permutation} labels a positroid configuration $C\!\in\!G(k,n)$, then \var{necklace} returns a list of $n$, $k$-tuples $\widehat{A}^{(a)}\equiv(\widehat{A}^{(a)}_1,\ldots,\widehat{A}^{(a)}_k)$ denoting the lexicographically-{\it maximal} non-vanishing minors starting from each column $a$. Like the more familiar `Grassmannian necklace' generated by the function \fun{necklace}, this data similarly encodes the ranks of all consecutive chains of columns of $C$; to see this, consider the `reverse' necklace for the configuration in $G(4,8)$ labeled by the permutation ${\color{perm}\{3,7,6,10,9,8,13,12\}}$,
\vspace{-.2cm}\eq{\begin{array}{c@{$\!\!\!$}c}\raisebox{-70.5pt}{\includegraphics[scale=.9]{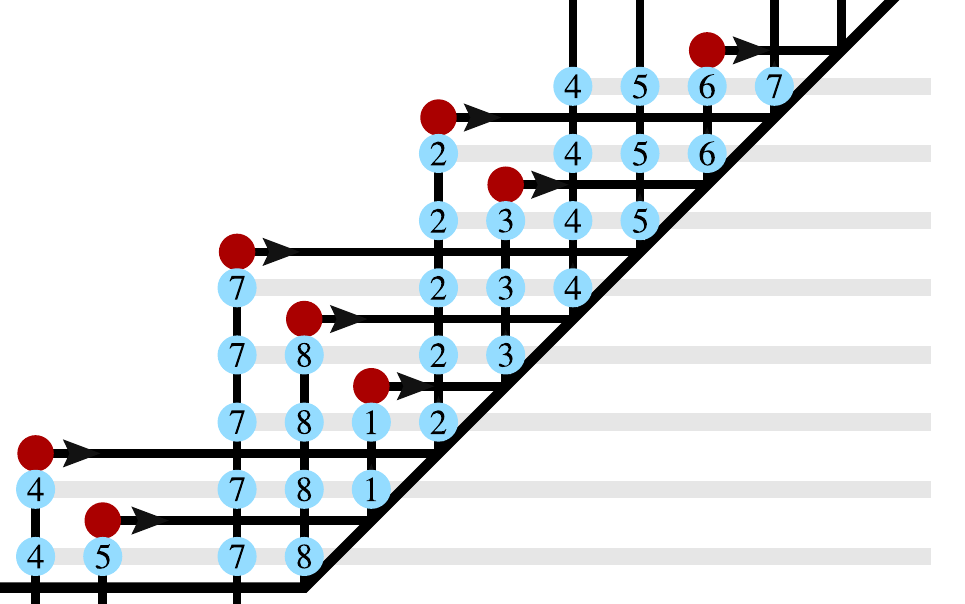}}&\begin{array}{l}~\\[-14.5pt]\widehat{A}^{(8)}=(4\,5\,6\,7)\\[0.8pt]\widehat{A}^{(7)}=(2\,4\,5\,6)\\[0.8pt]\widehat{A}^{(6)}=(2\,3\,4\,5)\\[0.8pt]\widehat{A}^{(5)}=(7\,2\,3\,4)\\[0.8pt]\widehat{A}^{(4)}=(7\,8\,2\,3)\\[0.8pt]\widehat{A}^{(3)}=(7\,8\,1\,2)\\[0.8pt]\widehat{A}^{(2)}=(4\,7\,8\,1)\\[0.8pt]\widehat{A}^{(1)}=(4\,5\,7\,8)\end{array}\\[10pt]\end{array}\vspace{-.2cm}}
This data is generated by the function \fun{necklaceR} according to:\\[5pt]
\mathematica{0.8}{necklaceR[\{3,7,6,10,9,8,13,12\}]\hspace{0pt}$~$}{{\tt\{\{4,5,7,8\},\{1,4,7,8\},\{1,2,7,8\},\{2,3,7,8\},\hspace{14cm}\phantom{1}\{2,3,4,7\},\{2,3,4,5\},\{2,4,5,6\},\{4,5,6,7\}\}}}
}

\defn{nonSingularQ}{\vardef{permutation},\vardefo{twistorDimension}{\tt 4}}{tests whether or not a configuration labeled by \var{permutation} has non-vanishing support for generic kinematical data---\fun{nonSingularQ} returns {\tt True} iff \mbox{\fun{intersectionNumber}[\var{permutation}]$>0$}.
}

\defn{parityConjugate}{\vardef{permutation}}{if \var{permutation} labels a configuration $C\!\in\!G(k,n)$, then \fun{parityConjugate} returns the permutation labeling the geometrically dual configuration $C^\perp\!\in\! G(n\,\mi\,k,n)$. 
For example,\\[5pt]
\mathematica{0.8}{parityConjugate[\{3,5,6,7,8,10\}]\hspace{14cm}$~$parityConjugate@\%}{{\tt \{4,5,7,6,8,9\}} ~\hspace{14cm}$~$ {\tt \{3,5,6,7,8,10\}}}
}

\defn{permToGeometry}{\vardef{permutation},\vardefo{removeVanishingQ}{\tt True}}{returns a (formatted) table of planes of various ranks spanned by consecutive chains of column-vectors of the configuration labeled by \var{permutation}; {\it distinguished} planes are highlighted in blue, and (for visual clarity) the option \var{removeVanishingQ} ({\tt True} by default) causes any vanishing columns of the configuration to be suppressed.

\ind For example, consider the positroid configuration labeled by the permutation ${\color{perm}\{3,7,6,10,9,8,13,12\}}$,
\eq{\hspace{-2.75cm}\begin{array}{c}\raisebox{-50pt}{\includegraphics[scale=1]{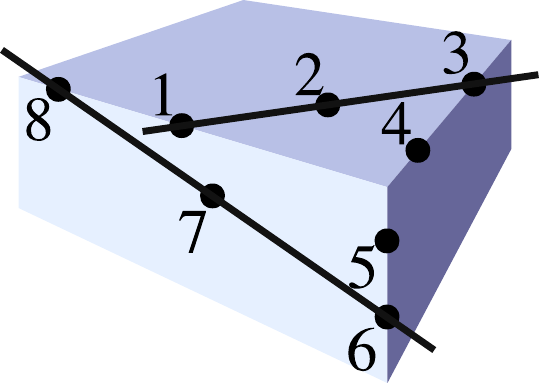}}\end{array}\,\quad\begin{array}{c|c}\multicolumn{2}{c}{}\\[-22pt]\text{consec. chains of columns}&\text{span}\\\hline{\color{deemph}(1)}\,(2)\,{\color{deemph}(3)\,(4)\,(5)\,(6)}\,(7)\,{\color{deemph}(8)}&\mathbb{P}^0\\(123)\,{\color{deemph}(34)\,(45)\,(56)}\,(678)\,{\color{deemph}(81)}&\mathbb{P}^1\\(56781)\,(81234)\,(3456)&\mathbb{P}^2\end{array}\vspace{-.05cm}\hspace{-2.0cm}\label{g48_configuration_example}}
For this, the function \fun{permToGeometry} would produce:\\[5pt]
\mathematica{0.8}{permToGeometry[\{3,7,6,10,9,8,13,12\}]\hspace{0pt}$~$}{\raisebox{-16pt}{$\left(\begin{array}{@{}c@{}}(1)\,({\color{blue}2})\,(3)\,(4)\,(5)\,(6)\,({\color{blue}7})\,(8)\\({\color{blue}1\,2\,3})(3\,4)({\color{blue}4\,5})\,(5\,6)\,({\color{blue}6\,7\,8})\,(8\,1)\\({\color{blue}3\,4\,5\,6})\,({\color{blue}5\,6\,7\,8\,1})\,({\color{blue}8\,1\,2\,3\,4})\end{array}\right)$}}\\[-24pt]

(Here, all the `distinguished' maximal planes $(a\pl1,\ldots,\sigma(a)\mi1)$ have been highlighted in blue---as the rest of the table follows form knowledge of these planes.)
}

\defn{permutationK}{\vardef{permutation}}{returns the ``$k$'' associated with \var{permutation}. More explicitly, if $C\equiv(c_1,\ldots,c_n)$ is the positroid configuration labeled by \var{permutation}, then \fun{permutationK} returns $\mathrm{rank}\{c_1,\ldots,c_n\}$; as such, $C\!\in\!G(k,n)$.
E.g.,\\[5pt]
\mathematica{0.8}{permutationK[\{3,7,6,10,9,8,13,12\}]}{{\tt 4}}
}

\defn{preferredGauge}{\vardef{permutation}}{returns the lexicographically first non-vanishing minor of the configuration $C\!\in\!G(k,n)$; equivalently, if $C\equiv(c_1,\ldots,c_n)$ is the configuration labeled by \var{permutation}, then \fun{preferredGauge} returns the lexicographically-minimal set of column-labels $(a_1,\ldots,a_k)$ such that $\mathrm{rank}\{c_{a_1},\ldots,c_{a_k}\}=k$. 
}

\defn{randomCell}{\vardef{n},\vardef{k},\vardef{d},\vardefo{exclusionsQ}{\tt True}}{gives the permutation label of a randomly-generated, \var{d}-dimensional positroid cell $C\!\in\!G($\var{k},\var{n}$)$. If the optional argument \var{exclusionsQ} is {\tt True} (its default value), then \fun{randomCell} tries to find a configuration for which all columns are non-vanishing (when possible).
}

\defntb{rotate}{\vardef{rotation}}{\vardef{permutation}}{returns the permutation labeling a positroid cell whose columns are rotated (positively) relative to that of \var{permutation} by \var{rotation}; more specifically, given a positroid cell $C=(c_1,\ldots,c_n)\!\in\!G(k,n)$ labeled by \var{permutation}, \fun{rotate} returns the label of the configuration $C'=(c_{1'},\ldots,c_{n'})$ where $a'=a\pl$\var{rotation}.
}

\defnNA{storeBoundaries}{}{a global variable which by default is set to {\tt False}; when {\tt True}, boundary information is stored in memory so that boundary computations need not be repeated. 
}

\newpage
\subsection{Positroid Coordinates and Matrix Representatives}

\defn{bridgeToMinors}{\vardef{permutation}}{using the BCFW-bridge construction of coordinates for the configuration $C$ labeled by \var{permutation}, \fun{bridgeToMinors} expresses each BCFW-bridge coordinate $\alpha_i$ directly in terms of the minors of $C$. For example, the ({\it canonical}) BCFW-bridge decomposition of the permutation ${\color{perm}\{3,7,6,10,9,8,13,12\}}$ generates the matrix representative:\\[5pt]
\mathematica{0.9}{permToMatrix[\{3,7,6,10,9,8,13,12\}]//nice\hspace{0pt}$~$}{\raisebox{-20pt}{\text{{\small$\displaystyle\left(\begin{array}{@{}cccccccc@{}}1&{\color[rgb]{0,0,0}\alpha_{9}}&0&\!\!{\color{black}\mi}\,{\color[rgb]{0,0,0}\alpha_{5}}\,&\!\!{\color{black}\mi}\,{\color[rgb]{0,0,0}\alpha_{5}} {\color[rgb]{0,0,0}\alpha_{6}}\,&0&0&0\\[-2pt]
0&1&{\color[rgb]{0,0,0}\alpha_{8}}&{\color[rgb]{0,0,0}\alpha_{7}}&0&0&0&0\\[-2pt]
0&0&0&1&{\color[rgb]{0,0,0}\alpha_{3}}\pl\,{\color[rgb]{0,0,0}\alpha_{6}}&{\color[rgb]{0,0,0}\alpha_{3}} {\color[rgb]{0,0,0}\alpha_{4}}&0&\!\!{\color{black}\mi}\,{\color[rgb]{0,0,0}\alpha_{1}}\\[-2pt]
0&0&0&0&1&{\color[rgb]{0,0,0}\alpha_{4}}&{\color[rgb]{0,0,0}\alpha_{2}}&0\end{array}\right)$}}}}

\vspace{-.75cm}It is a highly non-trivial fact that such bridge coordinates $\alpha_i$ can be expressed as ratios of {\it monomials} involving the minors of $C$. The explicit correspondence is given by the function \fun{bridgeToMinors}, as in the following example:\\[5pt]
\mathematica{0.9}{bridgeToMinors[\{3,7,6,10,9,8,13,12\}]//nice\hspace{0pt}$~$}{\raisebox{-16pt}{$\begin{array}{@{}llll@{}}\alpha_1\to\frac{(1\,2\,5\,8)}{(1\,2\,4\,5)}&\alpha_2\to\frac{(1\,2\,4\,7)}{(1\,2\,4\,5)}&\alpha_3\to\frac{(1\,2\,4\,5)(1\,2\,7\,8)(4\,5\,6\,7)}{(1\,2\,4\,6)(1\,2\,4\,7)(4\,5\,7\,8)}\\\alpha_4\to\frac{(1\,2\,4\,6)(1\,2\,6\,7)(4\,5\,7\,8)}{(1\,2\,4\,5)(1\,2\,7\,8)(4\,5\,6\,7)}&\alpha_5\to\frac{(1\,2\,4\,6)(4\,5\,7\,8)}{(1\,2\,7\,8)(1\,4\,5\,6)}&\alpha_6\to\frac{(1\,4\,5\,6)(2\,5\,7\,8)}{(1\,2\,4\,6)(4\,5\,7\,8)}\\\alpha_7\to\frac{(1\,4\,7\,8)}{(1\,2\,7\,8)}&\alpha_8\to\frac{(1\,3\,4\,5)}{(1\,2\,4\,5)}&\alpha_9\to\frac{(2\,4\,5\,6)}{(1\,4\,5\,6)}\end{array}$}}

}

\defn{matrixCharts}{\vardef{permutation}}{gives a list of (distinct) matrix-representatives of the positroid configuration $C$ labeled by \var{permutation}, using each of the (cyclically-distinct) columns as the `minimal' column used in the construction of BCFW-bridge coordinates for $C$. 
}

\defn{matrixToPerm}{\vardef{matrix}}{returns the permutation which labels the positroid cell represented by \var{matrix}. If \var{matrix} is given in terms of unspecified variables, then \fun{matrixToPerm} assumes that all such take on {\it generic} values.
}

\defn{permToMatrix}{\vardef{permutation},\vardefo{transpositionScheme}{\tt 0}}{returns a matrix-representative of the positroid cell labeled by \var{permutation} given in terms of BCFW-bridge coordinates obtained using one of several possible bridge-decomposition schema---with the default scheme being ``0'', that corresponding to the {\it canonical} or `lexicographic' scheme described in reference \cite{ArkaniHamed:2012nw}. The possible decomposition schema are those described for \fun{transpositionChain}. 
}

\defn{positiveQ}{\vardef{matrix}}{returns {\tt True} if the {\it all} ordered minors of the matrix are strictly non-negative; if \var{matrix} is parameterized by unspecified variables, then \fun{positiveQ} returns {\tt True} if \var{matrix} is positive (in the previous sense) when all its unfixed variables are given random, {\it positive} values.
}

\defn{transpositionChain}{\vardef{permutation},\vardefo{scheme}{\tt 0}}{returns a list {\tt \{transpositionList, permutationList, seedGauge\}} containing the complete BCFW-bridge decomposition of \var{permutation} into `adjacent' transpositions according to the scheme denoted \var{scheme}. (This data is obviously redundant, but such redundancy proves somewhat useful to have at hand.) The possible schema include:
\vspace{-.2cm}\eq{\begin{array}{|l|p{12cm}|}\hline\multicolumn{1}{|c|}{\var{scheme}}&\multicolumn{1}{c|}{\text{}}\\\hline\phantom{-}{\tt 0} \;\;\text{(default)}&the canonical or `lexicographic' decomposition scheme\\\hline\text{{\tt``cyclic''}}&a scheme which attempts to decompose \var{permutation} into a sequence of bridges in a way which preserves cyclic symmetry (if any exists)\\\hline\phantom{-}{\tt 1}&a scheme which {\it correctly orients} all $(2n\mi4)$-dimensional cells in the momentum-space Grassmannian\\\hline{\tt -1}&a scheme which {\it correctly orients} all $4k$-dimensional cells in the momentum-twistor Grassmannian\\\hline\end{array}\vspace{-.2cm}\nonumber}

\ind To illustrate the different transposition schemes, the {\it lexicographic} scheme (``\var{scheme}={\tt 0}'') would decompose the permutation ${\color{perm}\{4,7,6,9,8,11,10,13\}}$,\\[5pt]
\mbox{\hspace{-0.77cm}\begin{minipage}[h]{\textwidth}\eq{\hspace{-1.75cm}\begin{array}{|c@{$\,$}|@{$\,$}cccccccc@{$\,$}|c|}\multicolumn{10}{c}{\text{``0'' (Lexicographic) Bridge Decomposition Scheme}}\\\cline{1-10}\multicolumn{1}{|c@{$\,$}|@{$\,$}}{}&1&2&3&4&5&6&7&8&
\\[-4pt]\multicolumn{1}{|c@{$\,$}|@{$\,$}}{\tau}&\,\,\downarrow\,\,&\,\,\downarrow\,\,&\,\,\downarrow\,\,&\,\,\downarrow\,\,&\,\,\downarrow\,\,&\,\,\downarrow\,\,&\,\,\downarrow\,\,&\,\,\downarrow\,\,&\text{BCFW shift}
\\\cline{1-10}\multirow{2}{*}{({\color[rgb]{0.2,0,1.00}1\,2})}&{\color[rgb]{0.2,0,1.00}4}&{\color[rgb]{0.2,0,1.00}7}&6&9&8&11&10&13&\multirow{2}{*}{$c_{2}\mapsto c_{2}+{\color[rgb]{0.2,0,1.00}\alpha_{12}}c_{1}$}
\\\multirow{2}{*}{({\color[rgb]{0.188235,0,0.941}2\,3})}&7&{\color[rgb]{0.188235,0,0.941}4}&{\color[rgb]{0.188235,0,0.941}6}&9&8&11&10&13&\multirow{2}{*}{$c_{3}\mapsto c_{3}+{\color[rgb]{0.188235,0,0.941}\alpha_{11}}c_{2}$}
\\\multirow{2}{*}{({\color[rgb]{0.176471,0,0.882}3\,4})}&7&6&{\color[rgb]{0.176471,0,0.882}4}&{\color[rgb]{0.176471,0,0.882}9}&8&11&10&13&\multirow{2}{*}{$c_{4}\mapsto c_{4}+{\color[rgb]{0.176471,0,0.882}\alpha_{10}}c_{3}$}
\\\multirow{2}{*}{({\color[rgb]{0.164706,0,0.824}2\,3})}&7&{\color[rgb]{0.164706,0,0.824}6}&{\color[rgb]{0.164706,0,0.824}9}&{\color{deemph}4}&8&11&10&13&\multirow{2}{*}{$c_{3}\mapsto c_{3}+{\color[rgb]{0.164706,0,0.824}\alpha_{9\phantom{1}}}c_{2}$}
\\\multirow{2}{*}{({\color[rgb]{0.152941,0,0.765}1\,2})}&{\color[rgb]{0.152941,0,0.765}7}&{\color[rgb]{0.152941,0,0.765}9}&6&{\color{deemph}4}&8&11&10&13&\multirow{2}{*}{$c_{2}\mapsto c_{2}+{\color[rgb]{0.152941,0,0.765}\alpha_{8\phantom{1}}}c_{1}$}
\\\multirow{2}{*}{({\color[rgb]{0.141176,0,0.706}3\,5})}&{\color{deemph}9}&7&{\color[rgb]{0.141176,0,0.706}6}&{\color{deemph}4}&{\color[rgb]{0.141176,0,0.706}8}&11&10&13&\multirow{2}{*}{$c_{5}\mapsto c_{5}+{\color[rgb]{0.141176,0,0.706}\alpha_{7\phantom{1}}}c_{3}$}
\\\multirow{2}{*}{({\color[rgb]{0.129412,0,0.647}2\,3})}&{\color{deemph}9}&{\color[rgb]{0.129412,0,0.647}7}&{\color[rgb]{0.129412,0,0.647}8}&{\color{deemph}4}&6&11&10&13&\multirow{2}{*}{$c_{3}\mapsto c_{3}+{\color[rgb]{0.129412,0,0.647}\alpha_{6\phantom{1}}}c_{2}$}
\\\multirow{2}{*}{({\color[rgb]{0.117647,0,0.588}5\,6})}&{\color{deemph}9}&8&7&{\color{deemph}4}&{\color[rgb]{0.117647,0,0.588}6}&{\color[rgb]{0.117647,0,0.588}11}&10&13&\multirow{2}{*}{$c_{6}\mapsto c_{6}+{\color[rgb]{0.117647,0,0.588}\alpha_{5\phantom{1}}}c_{5}$}
\\\multirow{2}{*}{({\color[rgb]{0.105882,0,0.529}3\,5})}&{\color{deemph}9}&8&{\color[rgb]{0.105882,0,0.529}7}&{\color{deemph}4}&{\color[rgb]{0.105882,0,0.529}11}&{\color{deemph}6}&10&13&\multirow{2}{*}{$c_{5}\mapsto c_{5}+{\color[rgb]{0.105882,0,0.529}\alpha_{4\phantom{1}}}c_{3}$}
\\\multirow{2}{*}{({\color[rgb]{0.0941176,0,0.471}5\,7})}&{\color{deemph}9}&8&{\color{deemph}11}&{\color{deemph}4}&{\color[rgb]{0.0941176,0,0.471}7}&{\color{deemph}6}&{\color[rgb]{0.0941176,0,0.471}10}&13&\multirow{2}{*}{$c_{7}\mapsto c_{7}+{\color[rgb]{0.0941176,0,0.471}\alpha_{3\phantom{1}}}c_{5}$}
\\\multirow{2}{*}{({\color[rgb]{0.0823529,0,0.412}2\,5})}&{\color{deemph}9}&{\color[rgb]{0.0823529,0,0.412}8}&{\color{deemph}11}&{\color{deemph}4}&{\color[rgb]{0.0823529,0,0.412}10}&{\color{deemph}6}&{\color{deemph}7}&13&\multirow{2}{*}{$c_{5}\mapsto c_{5}-{\color[rgb]{0.0823529,0,0.412}\alpha_{2\phantom{1}}}c_{2}$}
\\\multirow{2}{*}{({\color[rgb]{0.0705882,0,0.353}5\,8})}&{\color{deemph}9}&{\color{deemph}10}&{\color{deemph}11}&{\color{deemph}4}&{\color[rgb]{0.0705882,0,0.353}8}&{\color{deemph}6}&{\color{deemph}7}&{\color[rgb]{0.0705882,0,0.353}13}&\multirow{2}{*}{$c_{8}\mapsto c_{8}+{\color[rgb]{0.0705882,0,0.353}\alpha_{1\phantom{1}}}c_{5}$}
\\&{\color{deemph}9}&{\color{deemph}10}&{\color{deemph}11}&{\color{deemph}4}&{\color{deemph}13}&{\color{deemph}6}&{\color{deemph}7}&{\color{deemph}8}&
\\\cline{1-10}\end{array}\hspace{0.3cm}\begin{array}{@{}c@{}}\\\raisebox{-3.25cm}{\includegraphics[scale=1.15]{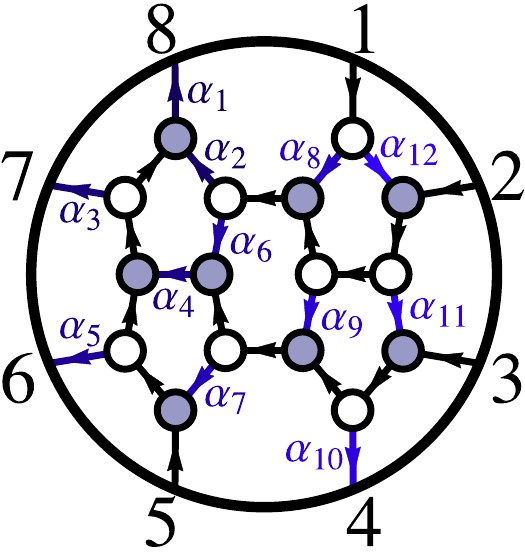}}\end{array}\hspace{-2cm}\nonumber}\end{minipage}}

This decomposition would give rise to the following representative of the cell:
{\small\eq{\left(\begin{array}{@{}cccccccc@{}}1&({\color[rgb]{0.152941,0,0.765}\alpha_{8}}\pl\,{\color[rgb]{0.2,0,1.00}\alpha_{12}})&({\color[rgb]{0.164706,0,0.824}\alpha_{9}}\pl\,{\color[rgb]{0.188235,0,0.941}\alpha_{11}}){\color[rgb]{0.152941,0,0.765}\alpha_{8}}&{\color[rgb]{0.152941,0,0.765}\alpha_{8}} {\color[rgb]{0.164706,0,0.824}\alpha_{9}} {\color[rgb]{0.176471,0,0.882}\alpha_{10}}&0&0&0&0\\
0&1&({\color[rgb]{0.129412,0,0.647}\alpha_{6}}\pl\,{\color[rgb]{0.164706,0,0.824}\alpha_{9}}\pl\,{\color[rgb]{0.188235,0,0.941}\alpha_{11}})&({\color[rgb]{0.129412,0,0.647}\alpha_{6}}\pl\,{\color[rgb]{0.164706,0,0.824}\alpha_{9}}){\color[rgb]{0.176471,0,0.882}\alpha_{10}}&({\color[rgb]{0.129412,0,0.647}\alpha_{6}} {\color[rgb]{0.141176,0,0.706}\alpha_{7}}\,\mi\,{\color[rgb]{0.0823529,0,0.412}\alpha_{2}})&\mi{\color[rgb]{0.0823529,0,0.412}\alpha_{2}} {\color[rgb]{0.117647,0,0.588}\alpha_{5}}&\mi{\color[rgb]{0.0823529,0,0.412}\alpha_{2}} {\color[rgb]{0.0941176,0,0.471}\alpha_{3}}&0\\
0&0&1&{\color[rgb]{0.176471,0,0.882}\alpha_{10}}&({\color[rgb]{0.105882,0,0.529}\alpha_{4}}\pl\,{\color[rgb]{0.141176,0,0.706}\alpha_{7}})&{\color[rgb]{0.105882,0,0.529}\alpha_{4}} {\color[rgb]{0.117647,0,0.588}\alpha_{5}}&0&0\\
0&0&0&0&1&{\color[rgb]{0.117647,0,0.588}\alpha_{5}}&{\color[rgb]{0.0941176,0,0.471}\alpha_{3}}&\!\!\mi\,{\color[rgb]{0.0705882,0,0.353}\alpha_{1}}\end{array}\right)}}

However, the ``{\tt cyclic}'' \var{scheme} would decompose the permutation according to:\\[5pt]
\mbox{\hspace{-0.77cm}\begin{minipage}[h]{\textwidth}\eq{\hspace{-1.75cm}\begin{array}{|c@{$\,$}|@{$\,$}cccccccc@{$\,$}|c|}\multicolumn{10}{c}{\text{``cyclic'' Bridge Decomposition Scheme}}\\\cline{1-10}\multicolumn{1}{|c@{$\,$}|@{$\,$}}{}&1&2&3&4&5&6&7&8&
\\[-4pt]\multicolumn{1}{|c@{$\,$}|@{$\,$}}{\tau}&\,\,\downarrow\,\,&\,\,\downarrow\,\,&\,\,\downarrow\,\,&\,\,\downarrow\,\,&\,\,\downarrow\,\,&\,\,\downarrow\,\,&\,\,\downarrow\,\,&\,\,\downarrow\,\,&\text{BCFW shift}
\\\cline{1-10}\multirow{2}{*}{({\color[rgb]{0.2,0,1.00}1\,2})}&{\color[rgb]{0.2,0,1.00}4}&{\color[rgb]{0.2,0,1.00}7}&6&9&8&11&10&13&\multirow{2}{*}{$c_{2\phantom{1}}\mapsto c_{2}+{\color[rgb]{0.2,0,1.00}\alpha_{12}}c_{1}$}
\\\multirow{2}{*}{({\color[rgb]{0.188235,0,0.941}3\,4})}&7&4&{\color[rgb]{0.188235,0,0.941}6}&{\color[rgb]{0.188235,0,0.941}9}&8&11&10&13&\multirow{2}{*}{$c_{4}\mapsto c_{4}+{\color[rgb]{0.188235,0,0.941}\alpha_{11}}c_{3}$}
\\\multirow{2}{*}{({\color[rgb]{0.176471,0,0.882}5\,6})}&7&4&9&6&{\color[rgb]{0.176471,0,0.882}8}&{\color[rgb]{0.176471,0,0.882}11}&10&13&\multirow{2}{*}{$c_{6}\mapsto c_{6}+{\color[rgb]{0.176471,0,0.882}\alpha_{10}}c_{5}$}
\\\multirow{2}{*}{({\color[rgb]{0.164706,0,0.824}7\,8})}&7&4&9&6&11&8&{\color[rgb]{0.164706,0,0.824}10}&{\color[rgb]{0.164706,0,0.824}13}&\multirow{2}{*}{$c_{8}\mapsto c_{8}+{\color[rgb]{0.164706,0,0.824}\alpha_{9\phantom{1}}}c_{7}$}
\\\multirow{2}{*}{({\color[rgb]{0.152941,0,0.765}2\,3})}&7&{\color[rgb]{0.152941,0,0.765}4}&{\color[rgb]{0.152941,0,0.765}9}&6&11&8&13&10&\multirow{2}{*}{$c_{3}\mapsto c_{3}+{\color[rgb]{0.152941,0,0.765}\alpha_{8\phantom{1}}}c_{2}$}
\\\multirow{2}{*}{({\color[rgb]{0.141176,0,0.706}3\,4})}&7&9&{\color[rgb]{0.141176,0,0.706}4}&{\color[rgb]{0.141176,0,0.706}6}&11&8&13&10&\multirow{2}{*}{$c_{4}\mapsto c_{4}+{\color[rgb]{0.141176,0,0.706}\alpha_{7\phantom{1}}}c_{3}$}
\\\multirow{2}{*}{({\color[rgb]{0.129412,0,0.647}6\,7})}&7&9&6&{\color{deemph}4}&11&{\color[rgb]{0.129412,0,0.647}8}&{\color[rgb]{0.129412,0,0.647}13}&10&\multirow{2}{*}{$c_{7}\mapsto c_{7}+{\color[rgb]{0.129412,0,0.647}\alpha_{6\phantom{1}}}c_{6}$}
\\\multirow{2}{*}{({\color[rgb]{0.117647,0,0.588}7\,8})}&7&9&6&{\color{deemph}4}&11&13&{\color[rgb]{0.117647,0,0.588}8}&{\color[rgb]{0.117647,0,0.588}10}&\multirow{2}{*}{$c_{8}\mapsto c_{8}+{\color[rgb]{0.117647,0,0.588}\alpha_{5\phantom{1}}}c_{7}$}
\\\multirow{2}{*}{({\color[rgb]{0.105882,0,0.529}1\,2})}&{\color[rgb]{0.105882,0,0.529}7}&{\color[rgb]{0.105882,0,0.529}9}&6&{\color{deemph}4}&11&13&10&{\color{deemph}8}&\multirow{2}{*}{$c_{2}\mapsto c_{2}+{\color[rgb]{0.105882,0,0.529}\alpha_{4\phantom{1}}}c_{1}$}
\\\multirow{2}{*}{({\color[rgb]{0.0941176,0,0.471}3\,5})}&{\color{deemph}9}&7&{\color[rgb]{0.0941176,0,0.471}6}&{\color{deemph}4}&{\color[rgb]{0.0941176,0,0.471}11}&13&10&{\color{deemph}8}&\multirow{2}{*}{$c_{5}\mapsto c_{5}+{\color[rgb]{0.0941176,0,0.471}\alpha_{3\phantom{1}}}c_{3}$}
\\\multirow{2}{*}{({\color[rgb]{0.0823529,0,0.412}5\,6})}&{\color{deemph}9}&7&{\color{deemph}11}&{\color{deemph}4}&{\color[rgb]{0.0823529,0,0.412}6}&{\color[rgb]{0.0823529,0,0.412}13}&10&{\color{deemph}8}&\multirow{2}{*}{$c_{6}\mapsto c_{6}+{\color[rgb]{0.0823529,0,0.412}\alpha_{2\phantom{1}}}c_{5}$}
\\\multirow{2}{*}{({\color[rgb]{0.0705882,0,0.353}7\,2})}&{\color{deemph}9}&{\color[rgb]{0.0705882,0,0.353}7}&{\color{deemph}11}&{\color{deemph}4}&{\color{deemph}13}&{\color{deemph}6}&{\color[rgb]{0.0705882,0,0.353}10}&{\color{deemph}8}&\multirow{2}{*}{$c_{2}\mapsto c_{2}-{\color[rgb]{0.0705882,0,0.353}\alpha_{1\phantom{1}}}c_{7}$}
\\&{\color{deemph}9}&{\color{deemph}2}&{\color{deemph}11}&{\color{deemph}4}&{\color{deemph}13}&{\color{deemph}6}&{\color{deemph}15}&{\color{deemph}8}&
\\\cline{1-10}\end{array}\hspace{0.3cm}\begin{array}{@{}c@{}}\\\raisebox{-3.25cm}{\includegraphics[scale=1.15]{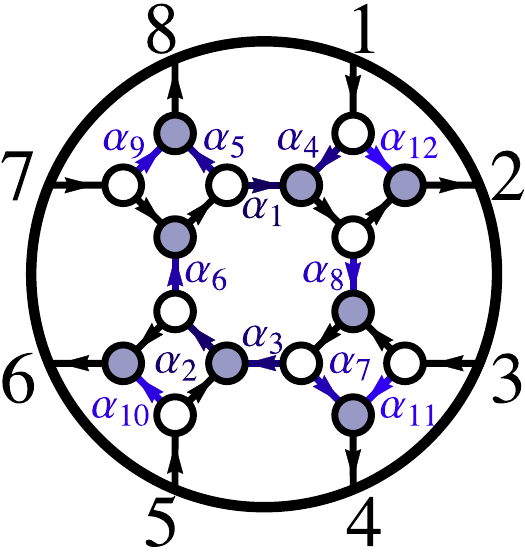}}\end{array}\hspace{-2cm}\nonumber}\end{minipage}}

which would result in the following matrix-representative:
{\small\eq{\left(\begin{array}{@{}cccccccc@{}}1&({\color[rgb]{0.105882,0,0.529}\alpha_{4}}\pl\,{\color[rgb]{0.2,0,1.00}\alpha_{12}})&{\color[rgb]{0.105882,0,0.529}\alpha_{4}} {\color[rgb]{0.152941,0,0.765}\alpha_{8}}&{\color[rgb]{0.105882,0,0.529}\alpha_{4}} {\color[rgb]{0.152941,0,0.765}\alpha_{8}} {\color[rgb]{0.188235,0,0.941}\alpha_{11}}&0&0&0&0\\
0&0&1&({\color[rgb]{0.141176,0,0.706}\alpha_{7}}\pl\,{\color[rgb]{0.188235,0,0.941}\alpha_{11}})&0&\!\!\mi\,{\color[rgb]{0.0823529,0,0.412}\alpha_{2}}&\!\!\mi\,{\color[rgb]{0.0823529,0,0.412}\alpha_{2}} {\color[rgb]{0.129412,0,0.647}\alpha_{6}}&\!\!\mi\,{\color[rgb]{0.0823529,0,0.412}\alpha_{2}} {\color[rgb]{0.129412,0,0.647}\alpha_{6}} {\color[rgb]{0.164706,0,0.824}\alpha_{9}}\\
0&0&0&0&1&({\color[rgb]{0.0941176,0,0.471}\alpha_{3}}\pl\,{\color[rgb]{0.176471,0,0.882}\alpha_{10}})&{\color[rgb]{0.0941176,0,0.471}\alpha_{3}} {\color[rgb]{0.129412,0,0.647}\alpha_{6}}&{\color[rgb]{0.0941176,0,0.471}\alpha_{3}} {\color[rgb]{0.129412,0,0.647}\alpha_{6}} {\color[rgb]{0.164706,0,0.824}\alpha_{9}}\\
0&{\color[rgb]{0.0705882,0,0.353}\alpha_{1}}&{\color[rgb]{0.0705882,0,0.353}\alpha_{1}} {\color[rgb]{0.152941,0,0.765}\alpha_{8}}&{\color[rgb]{0.0705882,0,0.353}\alpha_{1}} {\color[rgb]{0.152941,0,0.765}\alpha_{8}} {\color[rgb]{0.188235,0,0.941}\alpha_{11}}&0&0&1&({\color[rgb]{0.117647,0,0.588}\alpha_{5}}\pl\,{\color[rgb]{0.164706,0,0.824}\alpha_{9}})\end{array}\right)}}
}

\newpage
\subsection{Drawing On-Shell (Plabic) Graphs and Left-Right Paths}
\defn{plabicGraph}{\vardef{permutation},\vardefoo{optionsList}{\tt defaultOptions\!}}{draws a {\it reduced} plabic graph whose left-right path would be given by \var{permutation}. (Here, `plabic' is used somewhat loosely, as the default behavior of \fun{plabicGraph} is to draw graphs with monovalent and bivalent vertices when the inclusion of such is warranted.)

\ind There are many possible options for \fun{plabicGraph}; the principle of these are:\\[5pt]
\mbox{\hspace{-1.4cm}\begin{tabular}{|lc@{$\,$}|l|p{10cm}|}\multicolumn{1}{c}{option}&\multicolumn{1}{c}{}&\multicolumn{1}{c}{value}&\multicolumn{1}{c}{description}\\\hline\hline 
{\tt chainOption}&$\star$&{\tt \phantom{+}0}&uses the {\it lexicographic} bridge decomposition scheme\\\cline{3-4}&&{\tt \!`\!`cyclic'\!'\!}&uses the bridge decomposition scheme {\tt `\!`cyclic'\!'}\\\cline{3-4}&&$\!\!\pm${\tt 1}&uses the bridge decomposition scheme set by $\pm${\tt 1}\\\hline
{\tt rotation}&$\star$&{\tt \phantom{+}0}&constructs the graph using whatever bridge decomposition scheme is specified, but where particle `$1$' is considered minimal\\\cline{3-4}&&{\tt \phantom{+}a}&same as above, but with cyclic ordering beginning with particle $(1-${\tt a})\\\hline
{\tt LRpaths}&$\star$&{\tt \{\}}&does {\it not} show any left-right paths\\\cline{3-4}&&{\tt \!\{a,\ldots,b\}\!}&draws the left-right paths which start at legs {\tt a,\ldots,b}\\\cline{3-4}&&{\tt \!`\!`All'\!'\!}&draws {\it all} left-right paths\\\hline
{\tt directed}&$\star$&{\tt False}&results in an {\it undirected} graph\\\cline{3-4}&&{\tt True}&results in a {\it directed} graph whose {\it perfect orientation} follows from the BCFW bridge-decomposition scheme used\\\hline
{\tt edgeLabels}&$\star$&{\tt False}&does not label any edges of the graph\\\cline{3-4}&&{\tt True}&labels the BCFW-bridge edges by the bridge coordinates associated with each \\\hline
{\tt faceLabels}&$\star$&{\tt False}&does not label any faces of the graph\\\cline{3-4}&&{\tt True}&labels each face of the graph with a label ``$f_i$''\\\cline{3-4}&&{\tt \!`\!`A'\!'\!}&labels the faces of the graph with $A$-variable labels\\\cline{3-4}&&{\tt \!`\!`X'\!'\!}&labels the faces of the graph with $X$-variable labels\\\hline
{\tt showRemovable}&$\star$&{\tt True}&shows monovalent vertices attached to legs which are self-identified under \var{permutation}\\\cline{3-4}&&{\tt False}&removes boundary legs which are self-identified under \var{permutation}\\\hline
{\tt orientation}&$\star$&{\tt\phantom{-}1}&draws the external legs with {\it clockwise} ordering\\\cline{3-4}&&{\tt -1}&draws external legs with {\it counterclockwise} ordering\\\hline
{\tt bipartiteQ}&$\star$&{\tt False}&allows for trees of same-colored vertices\\\cline{3-4}&&{\tt True}&collapses all same-colored trees giving a bipartite graph\\\hline\multicolumn{3}{c}{}\\[-50pt]
\end{tabular}}
\newpage
Here, each {\bf default} option has been marked with a ``$\star$''. In addition to these options, there are many which control stylistic and aesthetic details of the graphs generated by \fun{plabicGraph}; the principle among these include: 
\\[5pt]
\mbox{\hspace{-1.59cm}\begin{tabular}{|lc@{$\,$}|l|p{10cm}|}\multicolumn{1}{c}{option}&\multicolumn{1}{c}{}&\multicolumn{1}{c}{value}&\multicolumn{1}{c}{description}\\\hline\hline 
{\tt angle}&$\star$&{\tt 0}&draws the graph with particle $1$ toward the middle-left\\\cline{3-4}&&{\tt $\theta$}&rotates the graph by {\tt $\theta$} radians in the {\it clockwise} direction\\\hline
{\tt extLabels}&$\star$&{\tt \!`\!`Auto'\!'\!}&labels the external legs {\tt \{1,\ldots,n\}}\\\cline{3-4}&&{\tt \!\{a,\ldots,b\}\!}&labels the external legs {\tt \{a,\ldots,b\}}\\\hline
{\tt labelSpacing}&$\star$&{\tt 0.65}&places the external particle labels at a distance of {\tt 0.65} from the boundary\\\hline
{\tt font}&$\star$&{\tt \!`\!`Times'\!'\!}&uses the font {\tt `\!`Times'\!'} for all labels\\\hline
{\tt fontSize}&$\star$&{\tt 36}&sets the ``{\tt FontSize}'' for external labels to be {\tt 36pt}\\\hline
{\tt imageSize}&$\star$&{\tt 300}&sets the ``{\tt ImageSize}'' of the \fun{Graphics} output to be {\tt 300}\\\hline
{\tt radius}&$\star$&{\tt 4}&sets the graph's boundary as a circle with radius {\tt 4} units\\\hline
{\tt vertexSize}&$\star$&{\tt 0.325}&draws all vertices with size of {\tt 0.325} units\\\hline
{\tt lineThickness}&$\star$&{\tt 3.5}&draws edges with \fun{AbsoluteThickness}{\tt[3.5]}\\\hline
{\tt LRpathThickness}&$\star$&{\tt 4}&draws left-right paths with \fun{AbsoluteThickness}{\tt[4]}\\\hline
{\tt LRpathDistance}&$\star$&{\tt 0.275}&draws left-right paths with a distance of {\tt 0.275} units from the graph's edges\\\hline
{\tt LRarrowHeadSize}&$\star$&{\tt 0.0655}&sets the size of left-right path \fun{Arrowheads} to be {\tt 0.0655}\\\hline
{\tt edgeArrowSize}&$\star$&{\tt 0.05}&sets the size of directed-edge \fun{Arrowheads} to be {\tt 0.05}\\\hline
{\tt outerCircle}&$\star$&{\tt True}&draws a disc at the boundary, enclosing the graph\\\hline
\end{tabular}}

Wherever sufficiently obvious, we have left as implicit the possible alternative settings allowed these options.

\ind For a complete list of options for \fun{plabicGraph}---together with the default value for each---consult the global variable ``\fun{defaultPlabicGraphOptions}''; users should find most detailed features of the output to be pliable through simple experimentation with the options.
}

\defn{plabicGraphData}{\vardef{permutation},\vardefo{transpositionScheme}{\tt 0}}{given the \var{permutation} label for a positroid configuration a some \var{transpositionScheme}, \fun{plabicGraphData} returns the pair {\tt \{edgeList,faceList\}}, where {\tt edgeList} lists each (directed) edge {\tt i}$\to${\tt j} (carrying a weight {\tt w}) as a triple {\tt \{i,j,w\}}; and {\tt faceList} lists the vertices along each face of the graph, listing them with clockwise ordering. 
}

\defnNA{resetGraphDefaults}{}{a function which automatically resets all the default graph options for the function \fun{plabicGraph}. (By default, \fun{plabicGraph} remembers any options explicitly set until the length of permutations being drawn is changed.)
}

\newpage
\subsection{Physical Operations, Kinematics, and Scattering Amplitudes}

\defn{bcfwPartitions}{\vardef{n},\vardef{k}}{returns a list of pairs $\{(n_L,k_L),(n_R,k_R)\}$ which are bridged-together to compute the \var{n}-point N$^{(\text{\var{k}}-2)}$MHV tree-amplitude $\mathcal{A}_\var{n}^{(\var{k})}$ according to \eq{\mathcal{A}_\var{n}^{(\var{k})}=\displaystyle\sum_{\substack{(n_L,k_L)\\(n_R,k_R)}}\mathcal{A}_{n_L}^{(k_L)}\underset{\mathrm{BCFW}}{\bigotimes}\mathcal{A}_{n_R}^{(k_R)}.\vspace{-.2cm}}
}

\defn{bcfwTermNames}{\vardef{n},\vardef{k}}{returns a {\it formatted} list of terms appearing in the BCFW tree-amplitude where MHV and $\bar{\text{MHV}}$ amplitudes have been collected together.\\[5pt]
\mathematica{0.95}{bcfwTermNames[6,3]\hspace{14cm}$~$
bcfwTermNames[14,7][[9598]]}{{\tt \{}$\mathcal{A}_5^{(3)}\!\!\otimes\!\mathcal{A}_3^{(1)},\mathcal{A}_4^{(2)}\!\!\otimes\!\mathcal{A}_4^{(2)},\mathcal{A}_3^{(2)}\!\!\otimes\!\mathcal{A}_5^{(2)}${\tt\}} ~\hspace{4cm}$~$ \text{{\normalsize$\big(\big(\mathcal{A}_3^{(2)}\!\!\otimes\!\big(\mathcal{A}_4^{(2)}\!\!\otimes\!\mathcal{A}_4^{(2)}\big)\big)\!\otimes\!\mathcal{A}_3^{(1)}\big)\!\otimes\!\big(\mathcal{A}_3^{(2)}\!\!\otimes\big(\big(\mathcal{A}_4^{(2)}\!\!\otimes\!\mathcal{A}_4^{(2)}\big)\!\otimes\!\mathcal{A}_3^{(1)}\big)\big)$}}}
}

\defntb{generalTreeContour}{\vardefo{a}{\tt 0},\vardefo{b}{\tt0},\vardefo{bridgeChoice}{\tt0}}{\vardef{n},\vardef{k}}{returns a list of permutation labels for the positroid cells which together give the \var{n}-particle N$^{(\var{k}-2)}$MHV tree-amplitude obtained using the {\it white-to-black} BCFW-bridge attached to legs $(1\,n)$ for \var{bridgeChoice} `{\tt 0}' (the default) or $(n\,1)$ for \var{bridgeChoice} `{\tt 1}', and for which the lower-point amplitudes appearing in the recursion have been recursed (using the same bridge-choice) attached to legs $(1\pl\,\var{a},n_L\pl\,\var{a})$ and $(1\pl\,\var{b},n_R\pl\,\var{b})$ of the left and right amplitudes, respectively. 
}

\defn{identitySigns}{\vardef{permList}}{returns a list of $\pm${\tt 1} of the relative signs needed for an identity involving the cells labeled by the permutations in \var{permList}. 
E.g.,\\[5pt]
\mathematica{0.95}{identitySigns[\{\{3,2,4,5,6,7\},\{2,4,3,5,6,7\},\{2,3,5,4,6,7\},\hspace{14cm}$~$\{2,3,4,6,5,7\},\{2,3,4,5,7,6\},\{1,3,4,5,6,8\}\}]}{{\tt \{1,-1,1,-1,1,-1\}}}
}

\defn{nRatioContour}{\vardef{n},\vardef{k}}{produces the same output as: 
\vspace{-.2cm}\eq{\text{\fun{permToResidue}{\tt /@}\fun{dualGrassmannian}{\tt /@}\fun{treeContour}{\tt [}\var{n},\var{k}{\tt]}}.\vspace{-.2cm}\nonumber}
}

\defn{nTreeContour}{\vardef{n},\vardef{k}}{produces the same output as:
\vspace{-.2cm}\eq{\text{\fun{permToResidue}{\tt /@}\fun{treeContour}{\tt [}\var{n},\var{k}{\tt]}}.\vspace{-.2cm}\nonumber}
}

\defn{permToResidue}{\vardef{permutation}}{for a \var{permutation} labeling either a $(2n\mi4)$-dimensional cell $C$ in the momentum-space Grassmannian or a $4k$-dimensional cell in the momentum-twistor Grassmannian, \fun{permToResidue} uses the globally defined momentum twistors {\tt Zs} and the corresponding (or alternatively-defined) global, spinor variables {\tt Ls} and {\tt Lbs} ($\lambda$ and $\widetilde{\lambda}$, respectively), to find the isolated point(s) $C^*\!\in\!C$ which solve the kinematical constraints and returns a pair (or list of pairs if more than one solution exists) $\{\mu^*,C^*\}$ where $\mu^*$ is the positroid measure evaluated at the point $C^*$ which solves the kinematical constraints.

\ind The on-shell function corresponding to a positroid cell labeled by \var{$\sigma$}, when evaluated at whatever kinematical data is given, would be given by,
\vspace{-.2cm}\eq{f_{\var{\sigma}}=\mu^*\!\!\times\!\delta^{k\times 4}\big(C^*\!\!\cdot\!\widetilde{\eta}),\quad\mathrm{or}\quad f_{\var{\sigma}}=\mu^*\!\!\times\!\delta^{k\times 4}\big(C^*\!\!\cdot\!\eta),\vspace{-.2cm}} 
if \var{$\sigma$} labels a cell in the momentum-space Grassmannian or the momentum-twistor Grassmannian, respectively. 
}

\defn{positiveZs}{\vardef{nParticles}}{it is sometimes convenient to evaluate analytic expressions involving spinor-helicity variables or momentum-twistors using explicit kinematical data; under such circumstances, there are some conveniences afforded by using ``well-chosen''  kinematical data.\\[-22pt]

\ind Reasons for preferring one choice over another include: having all Lorentz invariants be integer-valued and relatively small; having all dual-conformal cross ratios {\it positive} (so as to avoid branch-ambiguities when evaluating the polylogarithms that arise in scattering amplitudes at loop-level); and possibly to have all Lorentz-invariants be distinct (either to help reconstruct an analytic expression or to avoid `accidental' cancelations). Of these, the following momentum-twistors meet the first two desires spectacularly:
\vspace{-.2cm}\eq{{\tt Zs}\equiv\left(\begin{array}{@{}cccccc@{}}1&1&1&1&\cdots&\binom{n}{0}\\2&3&4&5&\cdots&\binom{n+1}{1}\\3&6&10&15&\cdots&\binom{n+2}{2}\\4&10&20&35&\cdots&\binom{n+3}{3}\end{array}\right).\vspace{-.2cm}}

\ind The function \fun{positiveZs}{\tt [16]} is evaluated when the {\tt positroids} package is first loaded, allowing amplitudes involving as many as $16$ particles to be evaluated without specific initialization. 
}

\defn{randomTreeContour}{\vardef{n},\vardef{k}}{returns the list of permutation labels for cells occurring in the BCFW tree-amplitude formula, where the lower-point amplitudes have been recursed using randomly-chosen legs. 
}

\defn{setupUsingSpinors}{\vardef{lambdaList},\vardef{lambdaBarList}}{sets up the global variables {\tt Ls} and {\tt Lbs} for $\lambda$ and $\widetilde{\lambda}$, respectively, and defines the global $(n\!\times\!4)$ matrix {\tt Zs} for momentum twistors for use in numerical evaluation. 
}

\defn{setupUsingTwistors}{\vardef{twistorList}}{sets up the global $(n\!\times\!4)$ matrix {\tt Zs} encoding the momentum-twistor kinematical data, and defines the auxiliary variables {\tt Ls} and {\tt Lbs} for $\lambda$ and $\widetilde{\lambda}$, respectively.
}

\defntb{superComponent}{\vardefms{component}}{\vardef{superFunction}}{for the purposes of the {\tt positroids} package, a \var{superFunction} must given by a pair $\{f,C\}$---an {\it ordinary} function $f(1,\ldots,n)$ of the kinematical variables times a {\it fermionic} $\delta$-function of the form,\\[-5pt]
\vspace{-.2cm}\eq{\delta^{k\times4}\big(C\!\cdot\!\widetilde{\eta}\big)\equiv\prod_{I=1}^4\left\{\bigoplus_{a_1<\!\cdots<a_k}\!\!(a_1\!\cdots a_k)\,\,\widetilde{\eta}_{a_1}^I\!\!\cdots\widetilde{\eta}_{a_k}^{I}\right\},\vspace{-.2cm}}
where $C$ is an $(n\times k)$-matrix of ordinary functions, and for each $a=1,\ldots,n$, $\widetilde{\eta}_a$ is a fermionic (anti-commuting) variable. To be clear, we consider each particle as a Grassmann coherent state of the form,
\vspace{-.2cm}\eq{\left|a \right> \equiv \left|a\right>_{\{\}} + \widetilde \eta_a^I  \left|a\right>_{\{I\}} + \frac{1}{2!} \widetilde \eta_a^I \widetilde \eta_a^J \left|a \right>_{\{I,J\}} + \frac{1}{3!} \widetilde \eta_a^I \widetilde \eta_a^J \widetilde \eta_a^K \left|a\right>_{\{I,J,K\}} + \widetilde \eta_a^1 \widetilde \eta_a^2 \widetilde \eta_a^3 \widetilde \eta_a^4 \left|a\right>_{\{1,2,3,4\}}; \nonumber\vspace{-.2cm}}
and if we use \var{$r_a$} to denote the $R$-charge of the $a^{\mathrm{th}}$ particle according to,

\begin{table}[h!]\begin{center}\vspace{-0.3cm}{\small\begin{tabular}{|l|l|l@{\hspace{1cm}}|l|}
\hline field &helicity&$R$-charge (\text{\var{$r_a$}})& short-hand for \var{$r_a$}\\\hline
$|a\rangle_{\{\}}$&$\;\;+1$&${\tt \{\}}$&$ {\tt p}$\\
$|a\rangle_{\{I\}}$&$\;\;+\frac{1}{2}$&${\tt \{I\}}$&${\tt p/2}(\Leftrightarrow{\tt \{4\}})$\\
$|a\rangle_{\{I,J\}}$&$\;\;\phantom{+}0$&{\tt \{I,J\}}&---\\
$|a\rangle_{\{I,J,K\}}$&$\;\;-\frac{1}{2}$&${\tt \{I,J,K\}}$&${\tt m/2}(\Leftrightarrow{\tt \{1,2,3\}})$\\
$|a\rangle_{\{1,2,3,4\}}$&$\;\;-1$&${\tt \{1,2,3,4\}}$&${\tt m}$\\\hline
\end{tabular}}\label{R_charges_intro}\end{center}\vspace{-0.85cm}
\end{table}
then \fun{superComponent}[\var{$r_1$},\ldots,\var{$r_n$}][\var{superFunction}] returns the {\it component} function of \var{superFunction} proportional to,
\vspace{-.4cm}\eq{\prod_{a=1}^{n}\prod_{I\in r_a}\widetilde{\eta}_a^{I}.\vspace{-.2cm}}
---that is, the component-function involving the states:
\vspace{-.2cm}\eq{|1\rangle_{\text{\var{$r_1$}}}\cdots |n\rangle_{\text{\var{$r_n$}}}.\vspace{-.3cm}}

\ind For example, the ``$(-,+,-,+,-,+,-,+)$'' component of the $8$-particle N$^2$MHV amplitude $\mathcal{A}_8^{(4)}$ proportional to,
\vspace{-.2cm}\eq{(\widetilde{\eta}_1^1\widetilde{\eta}_1^2\widetilde{\eta}_1^3\widetilde{\eta}_1^4)(\widetilde{\eta}_3^1\widetilde{\eta}_3^2\widetilde{\eta}_3^3\widetilde{\eta}_3^4)(\widetilde{\eta}_5^1\widetilde{\eta}_5^2\widetilde{\eta}_5^3\widetilde{\eta}_5^4)(\widetilde{\eta}_7^1\widetilde{\eta}_7^2\widetilde{\eta}_7^3\widetilde{\eta}_7^4),\vspace{-.2cm}}
would be extracted be obtained by evaluating, (compare with e.g.\ \cite{Dixon:2010ik,Bourjaily:2010wh}):\\[5pt]
\mathematica{0.8}{Total[superComponent[m,p,m,p,m,p,m,p]/@ \hspace{5cm} $~$ \hspace{.85cm} permToResidue/@treeContour[8,4]]}{\raisebox{-10pt}{$\displaystyle-\frac{\text{908}\,\text{416}}{\text{39}\,\text{375}}$}\vspace{0cm}}
\\[-40pt]}

\defn{termsInBCFW}{\vardef{n},\vardef{k},\vardefo{$\ell$}{\tt0}}{gives the number of (non-vanishing) terms generated by the BCFW-recursion for the \var{$\ell$}-loop, \var{n}-point N$^{(\var{k}-2)}$MHV amplitude. (\var{$\ell$} must be either {\tt 0} or {\tt 1} as the number of terms is scheme-dependent beyond 1-loop).\\[5pt]
\mathematica{0.8}{termsInBCFW[6,3] \hspace{14cm}$~$
termsInBCFW[6,3,1]}{{\tt 3}\hspace{14cm}$~$ {\tt 16}}
}

\defn{treeContour}{\vardef{n},\vardef{k}}{returns the list of permutation labels for positroid cells which together give the \var{n}-particle N$^{(\var{k}-2)}$MHV tree-amplitude, using the {\it default} recursion scheme; \fun{treeContour}{\tt [}\var{n},\var{k}{\tt]} is equivalent to \fun{generalTreeContour}{\tt[0,0,0][}\var{n},\var{k}{\tt]}.\\[5pt]
\mathematica{0.8}{treeContour[6,3]}{{\tt \{\{4,5,6,8, 7,9\},\{3,5,6,7,8,10\},\{4,6,5,7,8,9\}\}}}
}

\newpage
\subsection{Aesthetic and General Purpose Functions}

\defn{explicify}{\vardef{expression},\vardefo{positiveQ}{\tt True}}{picks random (integers) for all variables occurring in an \var{expression}; if \var{positiveQ} is {\tt True} (its default value), then the random assignments are taken to be positive. (If \var{expression} includes angle-brackets $\ab{\,\cdots}$, \fun{explicify} will evaluate these assuming random kinematical data.) 
}

\defntb{exportToPDF}{\vardef{fileName}}{\vardef{figure}}{saves a PDF version of \var{figure} to the file \var{filename}\footnote{The file is saved to the same directory as \fun{Export}{\tt[]}: either \fun{NotebookDirectory}{\tt []} or  \fun{Directory}{\tt[]} (if the former does not exist).} using outlined fonts (and with other minor processing).
}

\defn{mod2}{\vardef{objectList}}{returns the elements of \var{objectList} which occur an odd number of times in \var{objectList}. This can be useful, for example, if one wants to explicitly verify that $\partial^2=0\mod2$:\\[5pt]
\mathematica{0.9}{mod2[Join@@(boundary/@boundary[randomCell[8,4,12]])]}{{\tt \{\}}}
}

\defn{nice}{\vardef{expression}}{formats \var{expression} to display `nicely' by making replacements such as {\tt ab[x$\cdots$y]}$\mapsto\!\ab{x\cdots y}$, $\alpha${\tt [1]}$\mapsto\!\alpha_1$, etc., and by writing any level-zero matrices in {\tt MatrixForm}. 
}

\defn{niceTime}{\vardef{timeInSeconds}}{converts a time measured in seconds \var{timeInSeconds}, to human-readable form. For example,\\[5pt]
\mathematica{0.9}{niceTime[299\,792\,458]\hspace{14cm}$~$
niceTime[3.1415926535]}{{\tt 9 years, 182 days}\hspace{14cm}$~$ {\tt 3 seconds, 141 ms}}
}

\defn{random}{\vardef{objectList}}{returns a random element from (the first level of) \var{objectList}.
}

\defntb{randomSubset}{\vardef{subsetLength}}{\vardef{objectList}}{returns a randomly-chosen subset of length \var{subsetLength} from among the list \var{objectList}.
}

\defn{timed}{\vardef{expression}}{evaluates \var{expression} and prints a message regarding the time required for evaluation.
}

\newpage
\section*{Acknowledgements}
This work emerged directly out of research in collaboration with Nima Arkani-Hamed, Freddy Cachazo, Alexander Goncharov, Alexander Postnikov, and Jaroslav Trnka, and was greatly encouraged by early discussions with Pierre Deligne, Bob MacPherson, and Mark Goresky. We are especially grateful to Jaroslav Trnka, Nima Arkani-Hamed, and Freddy Cachazo for their helpful comments and suggestions regarding the package's documentation and the examples described in the demonstration notebook. We are indebted to Thomas Lam and David Speyer for their invaluable assistance in the development of a combinatorial test for kinematical support. This work was supported in part by the Harvard Society of Fellows, a grant form the Harvard Milton Fund, Department of Energy contract \mbox{DE-FG02-91ER40654}, and by the generous hospitality of the Institute for Advanced Study.

\newpage
%

\providecommand{\href}[2]{#2}\begingroup\raggedright\endgroup

~\newpage\thispagestyle{empty}
\thispagestyle{empty}

~\vspace{\fill}
\eq{\hspace{-5.0cm}\includegraphics[scale=1.4]{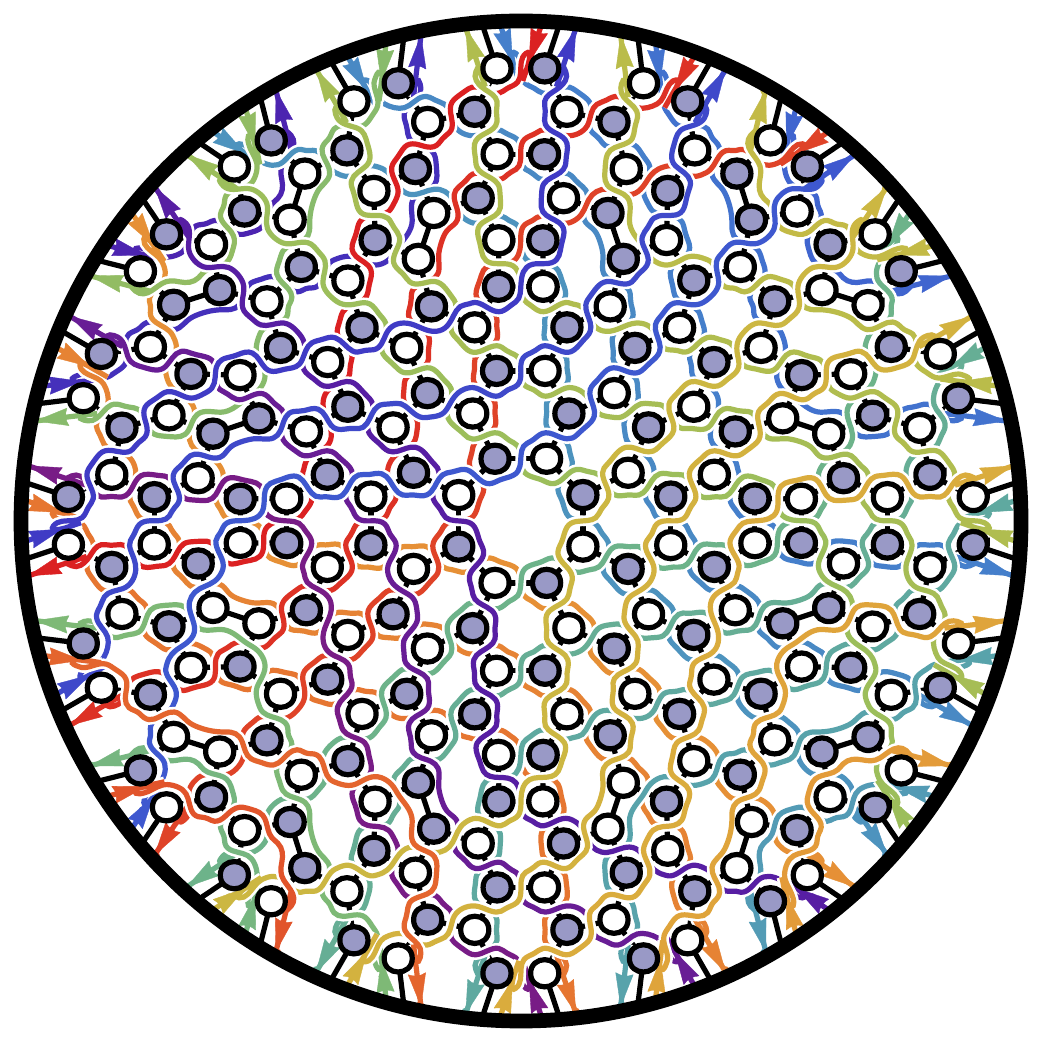}\hspace{-5cm}\nonumber}

\end{document}